\documentstyle[sprocl,epsf]{article}

\def\PLB{{\em Phys. Lett.}  B}

\def\be{\begin{equation}}
\def\ee{\end{equation}}
\def\bea{\begin{eqnarray}}
\def\eea{\end{eqnarray}}
\begin{document}

\title{$\bar{K}N$ INTERACTION AND KAONS IN THE NUCLEAR
MEDIUM}

\author{A. RAMOS}

 \address{Departament d'Estructura i Constituents de la
Mat\`eria,
Universitat de Barcelona, \\
Diagonal 647, 08028 Barcelona, Spain}

\author{E. OSET}

\address{
Departamento de
F\'{\i}sica Te\'orica and IFIC,
Universidad de Valencia-CSIC, \\
46100 Burjassot (Valencia), Spain}


\maketitle\abstracts{
The s-wave meson-nucleon interaction in the $S = -1$ sector is
studied by
means of a coupled-channel Lippmann Schwinger equation, using the
lowest order
chiral Lagrangian and a cut off to regularize the loop
integrals.
The position and width of the
$\Lambda (1405)$ resonance and the
$K^- p$
scattering cross sections at low energies
are well reproduced.
The inclusion
of the $\eta \Lambda, \eta \Sigma^0$ channels in the coupled
system is found
very important and allows a solution in terms of only the lowest
order Lagrangian.
The model is applied to calculate the in-medium $K^-$ self-energy
to which we add a small p-wave piece resulting from the
coupling to hyperon particle-nucleon hole excitations.
The $K^-$ feels an attraction of about $-100$ MeV
at normal nuclear density. The $\Lambda$(1405)
resonance shifts to energies above the $K^-p$ threshold and ends
up dissolving
as density increases. It
remains to be seen how these effects persist
when the dressing of the ${\bar K}$ and the $\pi$ mesons
is incorporated self-consistently in the calculation.
}

\section{Introduction}

Understanding the interaction of kaons ($K^+,K^0$) and antikaons
($\bar{K}^0,K^-$) with hadrons
is of especial interest for different branches of nuclear
physics.
From the more fundamental point of view, the fact that these
mesons
contain a strange quark, with a mass not so small
compared with the scale of chiral symmetry breaking, poses a
severe
test on the chiral Lagrangians. Moreover, due to the different
quark/antiquark
content, the $K^+=(u\bar{s})$ and the $K^-=(\bar{u}s)$ interact
very differently with nucleons.
The $K^-  p$ interaction, for instance,
is dominated
by the presence of a resonance, the $\Lambda(1405)$, which in a
quark
picture appears very naturally from the annihilation of the
$\bar{u}$
quark. Investigating whether the
chiral Lagrangians, written in terms of mesonic
and baryonic degrees of freedom, can account for
the differences between the $KN$ and
$\bar{K}N$ interactions is also a very challenging task.

The properties of the kaons and antikaons in the nuclear medium
have been the object of numerous investigations since the
possibility of the existence of a kaon
condensed phase in dense nuclear matter was pointed out
\cite{KN86}.
The presence of strangeness would influence the behaviour of
dense
matter due to the softening of the nuclear equation of state.
%leading to a smaller critical mass for gravitational collapse.
Most of the recent works start from chiral Lagrangians that
reproduce the free space scattering properties. The
$\Lambda(1405)$ resonance is either introduced as an elementary
field \cite{Lee9596} or dynamically generated through a
Lippmann-Schwinger equation \cite{Koch94,WKW96,Lutz98}.
Pauli blocking acting on the intermediate nucleon states makes
the
${\bar K}N$ interaction density dependent which, in turn,
modifies
the $K^-$ properties in the medium.
Medium effects at the simpler mean field level have been also
studied
in the context of chiral Lagrangians \cite{Li97} or with
relativistic Walecka-type models extended to incorporate
strangeness
in the form of hyperons or kaons \cite{Sch97}.

All the different approaches
agree qualitatively in establishing that, in
the
medium,
the $K^+$ feels a moderate repulsion and the $K^-$ a strong
attraction. How large is this attraction is still a subject under
intense debate.
Phenomenological approaches \cite{FGB94} based on fits to kaonic
atom data
find
a $K^-$-nucleus potential of
the order
of $-200 \pm 20$ MeV in the nuclear center. However,
no calculation that starts from the bare $K^- N$ interaction
predicts
such an attraction, the values
ranging from $-120$ to $-100$ MeV. Hopefully, heavy-ion
reactions, that are sensitive to higher density regions, will
help in elucidating these discrepancies. At the same time, it is
necessary to develop theories that treat the intricacies
associated to the mutual interaction between all the hadrons in
the medium as accurately as possible.

\section{$\bar{K}N$ interaction in free space}

The effective chiral Lagrangian formalism, which has been
very successful in
explaining the properties of meson-meson interaction at low
energies
\cite{Ga85,Pi95}, has also proved to be an excellent tool to
study the meson-baryon system
\cite{Eck95,Be95} when the interaction is weak, as in the case of
the
s-wave $\pi N$ and $K^+ N$ interaction.
However, in the $S = -1$ sector, the $\bar{K} N$ system
couples strongly
to many other channels and generates a resonance below threshold,
the $\Lambda (1405)$. In this case the standard chiral
scheme,
an expansion in powers of the typical momenta involved in the
process, fails
to be an appropriate approach.

A non perturbative scheme,
consisting of
%input of the chiral Lagrangians was employed in \cite{Kai95}. A
solving a set of
coupled-channel Lippmann Schwinger equations
using the lowest and next
to lowest order chiral Lagrangians, was employed in
Ref.~\cite{Kai95}.
The channels included
were $\bar{K}N$, $\pi \Lambda$ and $\pi \Sigma$ which are those
opened at the $K^- p$ threshold. Although a good
description of the $\Lambda (1405)$
resonance and the low energy
$K^-p$ cross sections
was obtained with only the lowest order Lagrangian,
the well measured threshold branching ratio,
$\gamma=\frac{\Gamma(K^- p
\to \pi^+ \Sigma^-)}{\Gamma(K^- p \to \pi^- \Sigma^+)}$, was
poorly
reproduced. This motivated the authors to include the $O(q^2)$
terms
of the chiral Lagrangian in the fit, at the
expense of introducing new parameters.

The success of this unitary coupled channel method
stimulated work in the meson-meson sector \cite{OO97} and
an excellent reproduction of the
resonances in the scalar sector, plus phase shifts and
inelasticities in the different physical channels was obtained,
by means
of coupled-channel Lippmann Schwinger equations using the lowest order
chiral Lagrangian,
plus a suitable cut off.

The model presented here extends the ideas of \cite{OO97} to
the
$\bar{K} N$ sector.
Although our approach shares many similarities with
that of Ref.~\cite{Kai95}, the main difference lies in
that our coupled equations also include
the channels
$\eta \Lambda$, $\eta \Sigma$ and $K \Xi$, which lie,
respectively, 230
MeV, 310 MeV and 380 MeV above the $K^-p$ threshold.
As we will see, the inclusion of these channels, especially the
$\eta \Lambda$, allows to describe all the low energy scattering
data,
including the ratio $\gamma$, with only the lowest order
Lagrangian.

\subsection{Meson-baryon amplitudes from the chiral Lagrangian}

The lowest order chiral
Lagrangian,
coupling the octet of pseudoscalar mesons to the octet of $1/2^+$
baryons, is
\begin{eqnarray}
L_1^{(B)} = &&\langle \bar{B} i \gamma^{\mu} \nabla_{\mu} B
\rangle -
M_B \langle \bar{B} B \rangle \nonumber \\
&&\frac{1}{2} D \langle \bar{B} \gamma^{\mu} \gamma_5 \left\{
u_{\mu},
B \right\} \rangle
+ \frac{1}{2} F \langle \bar{B} \gamma^{\mu} \gamma_5 [u_{\mu},
B]
\rangle \ ,
\end{eqnarray}
where the symbol $\langle\, \rangle$ denotes trace of SU(3)
matrices
and
\begin{equation}
\begin{array}{l}
\nabla_{\mu} B = \partial_{\mu} B + [\Gamma_{\mu}, B] \\
\Gamma_{\mu} = \frac{1}{2} (u^\dagger \partial_{\mu} u + u
\partial_{\mu} u^\dagger) \\
U = u^2 = {\rm exp} (i \sqrt{2} \Phi / f) \\
u_{\mu} = i u ^\dagger \partial_{\mu} U u^\dagger  \ .
\end{array}
\end{equation}
The SU(3) matrices for the mesons and the baryons are the
following
\begin{equation}
\Phi =
\left(
\begin{array}{ccc}
\frac{1}{\sqrt{2}} \pi^0 + \frac{1}{\sqrt{6}} \eta & \pi^+ & K^+
\\
\pi^- & - \frac{1}{\sqrt{2}} \pi^0 + \frac{1}{\sqrt{6}} \eta &
K^0 \\
K^- & \bar{K}^0 & - \frac{2}{\sqrt{6}} \eta
\end{array}
\right) \ ,
\end{equation}

\begin{equation}
B =
\left(
\begin{array}{ccc}
\frac{1}{\sqrt{2}} \Sigma^0 + \frac{1}{\sqrt{6}} \Lambda &
\Sigma^+ & p \\
\Sigma^- & - \frac{1}{\sqrt{2}} \Sigma^0 + \frac{1}{\sqrt{6}}
\Lambda & n \\
\Xi^- & \Xi^0 & - \frac{2}{\sqrt{6}} \Lambda
\end{array}
\right) \ .
\end{equation}
At lowest order in momentum the
interaction
Lagrangian reduces to
\begin{equation}
L_1^{(B)} = \langle \bar{B} i \gamma^{\mu} \frac{1}{4 f^2}
[(\Phi \partial_{\mu} \Phi - \partial_{\mu} \Phi \Phi) B
- B (\Phi \partial_{\mu} \Phi - \partial_{\mu} \Phi \Phi)]
\rangle     \ .
\end{equation}
The coupled channel formalism requires to evaluate the transition
amplitudes between the 10 different channels that can be built
from
the meson and baryon octets, namely $K^-p$, $\bar{K}^0 n$, $\pi^0
\Lambda$, $\pi^0 \Sigma^0$,
$\pi^+ \Sigma^-$, $\pi^- \Sigma^+$, $\eta \Lambda$, $\eta
\Sigma^0$,
$K^+ \Xi^-$ and $K^0 \Xi^0$.
These amplitudes have the form
\begin{equation}
V_{i j} = - C_{i j} \frac{1}{4 f^2} \bar{u} (p_i) \gamma^{\mu} u
(p_j)
(k_{i \mu} + k_{j \mu}) \ ,
\label{eq:v-matrix}
\end{equation}
where $p_j, p_i (k_j, k_i)$ are the initial, final momenta of the
baryons (mesons).
The explicit values of the coefficients $C_{ij}$ can be
found in ref. \cite{oset98}. At low energies we can neglect the
spatial components and Eq.~(\ref{eq:v-matrix}) simplifies to
\begin{equation}
V_{i j} = - C_{i j} \frac{1}{4 f^2} (k_j^0 + k_i^0) \ .
\end{equation}

\subsection{Coupled-channel Lippmann Schwinger equations}

The coupled-channel Lippmann Schwinger
equations in the center of mass frame read:
\begin{equation}
T_{i j} = V_{i j} + \overline{V_{i l} \; G_l \; T_{l j}}
\end{equation}
\noindent
where the indices $i,l,j$ run over all possible channels and
\begin{equation}
\overline{V_{i l} \; G_l \; T_{l j}} = i \int \frac{d^4 q}{(2
\pi)^4}
\, \frac{M_l}{E_l (\vec{q}\,)}
\, \frac{V_{i l} (k_i, q) \, T_{l j} (q, k_j)}{k^0 + p^0 - q^0 -
E_l
(\vec{q}\,)
+ i \epsilon} \, \frac{1}{q^2 - m^2_l + i \epsilon} \ ,
\label{eq:loop}
\end{equation}
with
$M_l, E_l$ and $m_l$ being the
baryon mass, baryon energy and meson mass in the intermediate
state.
The integral is regularized through a cut off, $q_{\rm max}$,
which  is the free parameter of our model.

Although Eq.~(\ref{eq:loop}) requires the half-off-shell
amplitudes, we can show, similarly as in Ref.
\cite{OO97}, that the off-shell part of the amplitude goes into
renormalization of coupling constants.
We take, as an example, the second order (one-loop) term in the
Lippmann Schwinger series
with equal
masses in the external and intermediate states for simplicity. We
have
\begin{equation}
\begin{array}{l}
V^2_{\rm off} = C (k^0 + q^0)^2 = C (2 k^0 + q^0 - k^0)^2 \\
\hspace{1cm}
= C^2 (2 k^0)^2 + 2 C (2 k^0) (q^0 - k^0) + C^2 (q^0 - k^0)^2 \ ,
\end{array}
\label{eq:onshell}
\end{equation}
with $C$ a constant. The first term in the last expression is the
 on shell
contribution $V^2_{\rm on} (V_{\rm on} \equiv C 2 k^0)$.
Neglecting
$p^0 - E (\vec{q}\,)$
in Eq. (\ref{eq:loop}), typical approximation in the heavy
baryon formalism,
the $(q^0-k^0)$ factor of the second term in Eq.
(\ref{eq:onshell}) cancels the baryon propagator.  The
contribution
is proportional to $V_{\rm on}$ and, therefore, is already
absorbed
by the first order (no loop) term
because the physical coupling $f$
is
used. Similarly,
the term proportional to $(q^0 - k^0)^2$ will cancel
the
$(k^0 - q^0)$ term in the denominator. The remaining factor,
$(k^0 - q^0)$, contains a term proportional to $k^0$
(and hence
$V_{\rm on}$) and a term proportional to $q^0$ which vanishes for
parity reasons.
Similar arguments can be used in the case of coupled channels and
higher order loops and the conclusion is that
we can factorize $V_{\rm on}$ and $T_{\rm on}$ outside the
integral of
Eq. (\ref{eq:loop}). As a consequence, the integral equation
reduces
to a set of algebraic equations and the loop integral reads
\begin{eqnarray}
G_{l} &=& i \, \int \frac{d^4 q}{(2 \pi)^4} \, \frac{M_l}{E_l
(\vec{q}\,)} \,
\frac{1}{k^0 + p^0 - q^0 - E_l (\vec{q}\,) + i \epsilon} \,
\frac{1}{q^2 - m^2_l + i \epsilon} \nonumber \\
&=& \int_{\mid {\vec q} \mid < q_{\rm max}} \, \frac{d^3 q}{(2
\pi)^3} \,
\frac{1}{2 \omega_l (\vec q\,)}
\,
\frac{M_l}{E_l (\vec{q}\,)} \,
\frac{1}{\sqrt{s}- \omega_l (\vec{q}\,) - E_l (\vec{q}\,) + i
\epsilon}
\label{eq:gprop}
\end{eqnarray}
with $\sqrt{s} = p^0 + k^0$.

Our model depends on a free parameter, $q_{\rm max}$, which is
determined
such that, for a fixed value
of $f$ chosen
between $f_{\pi} = 93$ MeV and $f_{K}=1.22 f_{\pi}$, we get the
best reproduction of the threshold ratios: $\gamma$, $R_c=
\frac{\Gamma (K^- p \rightarrow {\rm\scriptstyle charged\
particles})}{\Gamma (K^- p \rightarrow {\rm\scriptstyle all})}$ and
$R_n = \frac{\Gamma (K^- p \rightarrow \pi^0 \Lambda)}
{\Gamma (K^- p \rightarrow {\rm\scriptstyle all\ neutral\
states})}$.
The precise value of $f$
is the one that gives the best agreement with the
$\Lambda(1405)$ properties seen in the
$\pi \Sigma$ mass spectrum.
Our optimal choice was found for $f = 1.15 f_{\pi}$ and $q_{\rm
max} = 630$ MeV.

\subsection{$K^-p$ scattering observables}

The results for the threshold ratios
are shown in Table \ref{tab:ratios} where they are compared with
those obtained omitting the
$\eta$ channels.
We can see that the three ratios are reproduced within
$5 \%$. The ratio $\gamma$ is reduced by a factor
2.2 when the $\eta$ channels are omitted.
This small value for $\gamma$
is compatible with that obtained in
\cite{Kai95}, which motivated the authors
of that work to incorporate
higher order terms in the chiral expansion and perform a global
fit with five parameters.

\begin{table}[ht]
\caption{Threshold ratios\label{tab:ratios}}
\vspace{0.2cm}
\begin{center}
\footnotesize
\begin{tabular}{|l|c|c|c|}
\hline
{} &\raisebox{0pt}[13pt][7pt]{$\gamma$} &
\raisebox{0pt}[13pt][7pt]{$R_c$}
& \raisebox{0pt}[13pt][7pt]{$R_n$} \\
\hline
 \ & \ & \ & \ \\
All channels & 2.32 & 0.627 & 0.213 \\
No $\eta$ & 1.04 & 0.637  & 0.158 \\
exp. \cite{No78,To71}& 2.36 $\pm$ 0.04 & 0.664 $\pm$ 0.011 &
0.189
$\pm$
0.015 \\
[5pt]
\hline
\end{tabular}
\end{center}
\end{table}

\begin{figure}[ht]
       \setlength{\unitlength}{1mm}
       \begin{picture}(80,60)
       \put(25,0){\epsfxsize=9cm \epsfbox{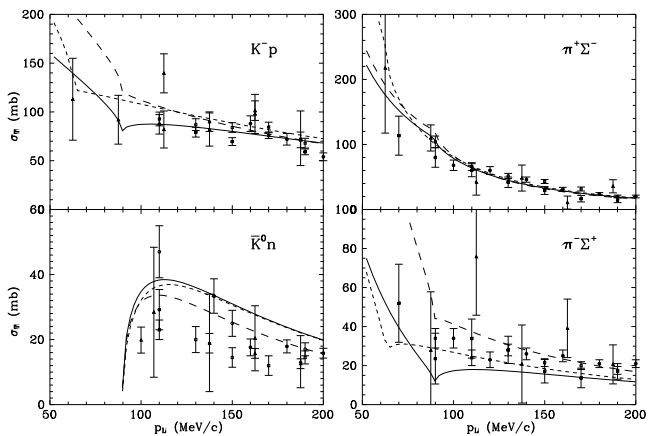}}
       \end{picture}
%%\rule{5cm}{0.2mm}\hfill\rule{5cm}{0.2mm}
%%\vskip 10 cm
%%\rule{5cm}{0.2mm}\hfill\rule{5cm}{0.2mm}
%\psfig{figure=mes98fig1.ps,height=1.5in}
\caption{
$K^-p$ scattering cross sections as functions of the $K^-$
momentum in the lab frame: with the full basis of physical states
(solid line), omitting the $\eta$ channels (long-dashed line) and
with the isospin-basis (short-dashed line).
\label{fig:mes98fig1}}
\end{figure}

The $K^-p$ cross sections in all channels, given by
\begin{equation}
\sigma_{i j} = \frac{1}{4 \pi} \, \frac{M_i M_j}{s} \,
\frac{k_i}{k_j} \, |T_{i j}|^2
\label{eq:xsec}
\end{equation}
with $j=1\ (K^- p)$,
are then calculated with the best choice of parameters and have
not been used in a best
fit to the data. Our results for some selected channels ($K^-p
\to K^- p,\bar{K}^0n,\pi^+\Sigma^-,\pi^-\Sigma^+$) are compared
with the low-energy scattering data in Fig.~\ref{fig:mes98fig1}.
We show the cross sections obtained with
the full basis of physical states (solid line), with the isospin
basis which uses average masses for $\bar{K},\pi,N$ and $\Sigma$
(short-dashed line), and omitting the coupling to the
intermediate $\eta$ channels (long-dashed line).
The results using the isospin basis are close to those using the
basis of physical states but the cusp associated to the
opening of the $\bar{K}^0 n$ channel appears in the wrong place.
The effects of the $\eta$ channels are much more
spectacular.
%, especially in the $\pi\Sigma$ cross sections.
Close to threshold, the $\pi^-\Sigma^+$ cross section is reduced
by almost a factor of 3 and
the $\pi^+ \Sigma^-$ cross section is reduced by a factor 1.3
when the
$\eta$ channels are included. This enhances the ratio $\gamma$ by
a factor 2.2 and makes the agreement with the experimental value
possible using only the lowest order chiral Lagrangian.

Finally, the $K^-p$ and $K^-n$ scattering lengths, defined as
\begin{equation}
a_i = -\frac{1}{4\pi} \frac{M}{\sqrt{s}} T_{ii} \hspace{3cm}
i=K^-p\ \ {\rm or}\ \ K^- n \ ,
\label{eq:length}
\end{equation}
 are presented
in Table~\ref{tab:length}.
We observe that isospin
breaking effects in the $K^- n$ amplitude, as well
as those omitting the $\eta$ channel, are moderate in this case.
We should
note that this is an isospin $I = 1$ channel where the
The $K^- p$ scattering length is essentially in
agreement
with the most recent results from Kaonic hydrogen $X$ rays
\cite{Miw97}.
%$(- 0.78 \pm 0.15 \pm 0.03) + i (0.49 \pm 0.25 \pm 0.12)$ fm.
Our results are also in
qualitative agreement with the scattering lengths determined from
scattering
data in \cite{Adm81} which have an estimated error of 15\%.
Note that these results are
obtained from the isospin scattering lengths
but as we can see from our calculations shown in
Table~\ref{tab:length} there are
violations of isospin at the
level of $20 \%$ in these amplitudes.
It is also worth calling the attention to the remarkable
agreement of
our results for the real part of the scattering lengths with
those
obtained in \cite{Adm81} from a combined dispersion relation and
$M$ matrix analysis.

\begin{table}[ht]
\caption{$K^- N$ scattering lengths \label{tab:length}}
\vspace{0.2cm}
\begin{center}
\footnotesize
\begin{tabular}{|l|c|c|}
\hline
{} &\raisebox{0pt}[13pt][7pt]{$a_{K^- p}$ (fm)} &
\raisebox{0pt}[13pt][7pt]{$a_{K^- n}$ (fm)} \\
\hline 
 \ & \ & \  \\
Full basis & $- 1.00 + i 0.94$  & $0.53 + i 0.62$  \\
No $\eta$ & $- 0.68 + i 1.64$   & $0.47 + i 0.53$ \\
Isos. basis & $- 0.85 + i 1.24$ & $0.54 + i 0.54$  \\
Exp. \cite{Miw97} & $(-0.78\pm 0.18) + i(0.49\pm 0.37)$ & \\
Exp. \cite{Adm81} & $- 0.67 + i 0.64$ & $0.37 + i 0.60$ \\
Exp. $Re\,(a)$\cite{Adm81} & $- 0.98 $ & $0.54$ \\
[5pt]
\hline
\end{tabular}
\end{center}
\end{table}

\section{$\bar{K}$ in the nuclear medium}

The dynamics of the ${\bar K} N$ interaction at low energies is
dominated by the $\Lambda(1405)$ resonance, an isospin $I=0$
quasi-bound
$K^- p$ state. As a result, the isospin averaged ${\bar K}N$
interaction is repulsive in free space.
However, kaonic atom data suggest that the $K^-$ feels a strongly
attractive potential at nuclear densities $\rho\sim \rho_0$,
where $\rho_0$ is normal nuclear matter density. This implies a
rapid
transition from a repulsive ${\bar K}N$ interaction to an
attractive one as density increases and, therefore, the study of
the
$K^-$ properties in the medium cannot be done in terms of the
simple $T\rho$ or impulse approximation. It is therefore
necessary to consider
the density dependence of the in-medium ${\bar K}N$
interaction, $T^{\rm eff}(\rho)$.

\subsection{Pauli blocking}

\begin{figure}[ht]
%\rule{5cm}{0.2mm}\hfill\rule{5cm}{0.2mm}
%\vskip 6 cm
%\rule{5cm}{0.2mm}\hfill\rule{5cm}{0.2mm}
%\psfig{figure=filename.ps,height=1.5in}
       \setlength{\unitlength}{1mm}
       \begin{picture}(80,60)
       \put(25,5){\epsfxsize=6cm \epsfbox{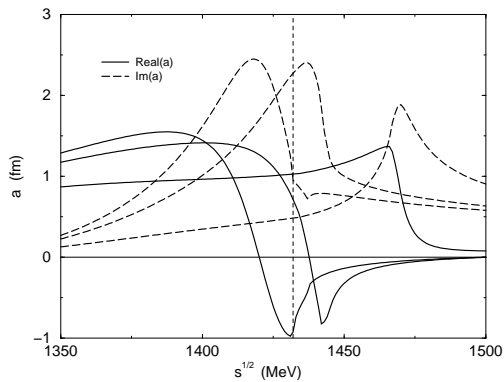}}
       \end{picture}
\caption{
Real (solid line) and Imaginary (dashed line) parts of the $K^-
p$ scattering length at three different Fermi momenta $p_F=0$,
150 and 300 MeV/c.
\label{fig:mes98fig2}}
\end{figure}

One source of density dependence comes from the
Pauli principle,
which prevents the scattering to intermediate nucleon states
below the
Fermi momentum, $p_F$, and ends up blocking the appearance of the
$\Lambda(1405)$ resonance. In Fig. \ref{fig:mes98fig2} we show
the
real (solid lines) and imaginary (dashed lines) parts of the
in-medium scattering length as functions of $\sqrt{s}$ for the
scattering of a $K^-$ with a proton at rest. The results are
shown for $p_F=0$, 150 and 300 MeV/c and, as density increases,
the $\Lambda(1405)$ resonance moves to higher values of
$\sqrt{s}$, hence changing from being slightly below
the $K^- p$ threshold (denoted by the vertical line) at $p_F=0$
MeV/c
to being
above it from a certain density on. It is also clearly seen that,
already at $p_F=150$ MeV/c, the scattering length at threshold is
positive, which  means that the in-medium $K^- p$ interaction has
become attractive.

\begin{figure}[ht]
%\rule{5cm}{0.2mm}\hfill\rule{5cm}{0.2mm}
%\vskip 5 cm
%\rule{5cm}{0.2mm}\hfill\rule{5cm}{0.2mm}
%\psfig{figure=filename.ps,height=1.5in}
       \setlength{\unitlength}{1mm}
       \begin{picture}(80,60)
       \put(25,5){\epsfxsize=6cm \epsfbox{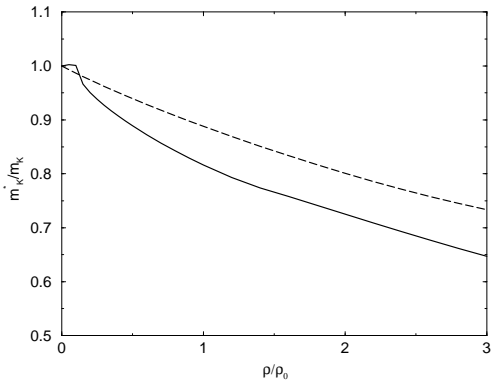}}
       \end{picture}
\caption{
Effective mass of the $K^-$ as a function of density in neutron
matter
(dashed line) and symmetric nuclear matter (solid line).
\label{fig:mes98fig3}}
\end{figure}

The $K^-$ properties in the medium can be derived from the $K^-$
self-energy (or optical potential) which is obtained by summing
up the in-medium
${\bar K}N$ interaction for all the nucleons filling up the Fermi
sea:
\begin{equation}
\Pi^s_{K}(q^0,{\vec q},\rho)=2\int \frac{d^3p}{(2\pi)^3}
\theta (p_F-\mid {\vec p} \mid ) \left[ T^{\rm eff}_{K^-
p}(\sqrt{s},\rho) +
T^{\rm eff}_{K^- n}(\sqrt{s},\rho) \right] \ .
\label{eq:selfka}
\end{equation}
We find that the averaged in-medium interaction defined as
$\overline{T}^{\rm eff}(\sqrt{s},\rho)=
\Pi^{s}_{K}(q^0=m_K,{\vec q}=0,\rho)/\rho$ becomes
attractive around $\rho \simeq 0.17 \rho_0$ ($p_F \simeq 150$
MeV/c). The solution of the new
$K^-$ dispersion relation
\begin{equation}
\omega^2 = \vec{q}\,^2 + m_K^2 + \Pi^{s}_{K}(\omega,{\vec
q},\rho) \ ,
\label{eq:disp}
\end{equation}
determines the effective mass, $m^*_K={\rm
Re}\, \omega({\vec q}=0)$, and decay width, $\Gamma=-2\,{\rm Im}
\,\omega({\vec q}=0)$, of the $K^-$ meson in the medium.
In Fig.~\ref{fig:mes98fig3} we show the $K^-$ effective mass in
units of
the bare mass as a function of the density for neutron matter
(dashed
line) and symmetric nuclear matter (solid line). The effective
mass
shows a strong non-linear density dependence at very small
densities in symmetric nuclear matter due to the influence of the
$\Lambda$(1405)
resonance in the $K^- p$ interaction. As density increases,
the $K^-$ pole lies far away from its free space value and the
effect
of the resonance, if still present, is negligible.
At $\rho_0$ we find
$m^*_K = 403$ MeV, implying an equivalent s-wave $K^-$-nucleus
potential $U_{K^-} = \Pi^{s}_{K}/2\omega \simeq -100$ MeV,
much smaller in magnitude than the value suggested by the
atomic data analyses. The
decay width of the $K^-$ is relatively small and amounts to
$\Gamma
= 34$ MeV. Our results for the in-medium ${\bar K}N$
interaction and the $K^-$ properties are in qualitative agreement
with
those of Ref.~\cite{WKW96},
where the chiral approach of \cite{Kai95} with the updated
parameters of \cite{Kai97} is used.

\subsection{Many-body correlations}

If Pauli blocking can affect so drastically the behaviour of
the $K^-$
in the medium with respect to that in free space it is reasonable
to investigate
other sources of density dependence which are, a priori, equally
important. In particular, all the mesons
and baryons participating in the intermediate loops interact with
the
nucleons of the Fermi sea and, as a consequence, feel an optical
potential which changes the threshold  energy of the different
channels.

For the
time being, we are going to assume that all the relevant baryons
(N,$\Lambda$,$\Sigma$) feel the same moderately attractive
potential,
which cancels out in the differences involved in the energy
balances.
We will instead focus our attention on the dressing of the
mesons,
particularly that of ${\bar K}$ and $\pi$ since the
$\eta$ appears in intermediate states that lie quite far above
the $K^- p$ threshold.

The model discussed here provides a s-wave $K^-$ self-energy
[Eq.~(\ref{eq:selfka})] to which we also add a p-wave
contribution
coming from the coupling of the $\bar{K}$ meson to
hyperon particle--nucleon hole excitations:
\begin{equation}
\Pi^{p}_{K}(q,\rho) = \frac{1}{2}
\left(\frac{g_{\scriptscriptstyle N\Lambda
K}}{2 M}\right)^2 {\vec q}\,^2 U_\Lambda(q,\rho) +
\frac{3}{2} \left(\frac{g_{\scriptscriptstyle N\Sigma K}}{2
M}\right)^2 {\vec q}\,^2 U_\Sigma(q,\rho)  \ ,
\label{eq:selfkap}
\end{equation}
where $g_{\scriptscriptstyle N\Lambda K},
g_{\scriptscriptstyle N\Sigma K}$ are the SU(3) strong coupling
constants and $U_Y(q,\rho)$ is the Lindhard function for a
hyperon
particle-nucleon hole excitation.
This p-wave contribution can become
important for large values of the $K^-$ momentum.

For the pion self-energy we take that of Ref.
\cite{ramos94} which consists of a small constant s-wave
part, $\Pi_\pi^s(\rho)$, plus a p-wave part,
$\Pi_\pi^p(q,\rho)$,
modified by the
effect of short range nuclear correlations through the
typical $g^\prime$ Landau parameter (slightly momentum
dependent):
\begin{equation}
\Pi^p_\pi(q,\rho)=\left(\frac{f}{m_\pi}\right)^2 {\vec q}\,^2
\frac{U_\pi(q,\rho)}{1-\left(\frac{f}{m_\pi}\right)^2 g^\prime(q)
U_\pi(q,\rho)} \ ,
\end{equation}
where $U_\pi(q,\rho) = U_N(q,\rho) + U_\Delta(q,\rho) + U_{\rm
2p2h}(q,\rho)$ contains the Lindhard functions for
particle-hole and $\Delta$-hole excitations plus a piece coming
from the coupling to 2 particle-2 hole excitations.

The dressed meson propagator reads ($i={\bar K},\pi$):
\begin{equation}
D_i(q,\rho) = \frac{1}{(q^0)^2-{\vec q\,}^2 - m_i^2 -
\Pi_i(q,\rho)}
= \int_0^\infty d\omega \, 2\omega  \, \frac{S_i(\omega,
{\vec q},\rho)}{(q^0)^2 - \omega^2 + i\epsilon}  \ ,
\label{eq:dressed}
\end{equation}
written in terms of the self-energy (second term) or the spectral
density (last term) which is defined as $S_i(\omega,{\vec
q},\rho)= - {\rm Im}\, D_i(\omega,{\vec q},\rho)/\pi $.

\begin{figure}[ht]
%\rule{5cm}{0.2mm}\hfill\rule{5cm}{0.2mm}
%\vskip 5 cm
%\rule{5cm}{0.2mm}\hfill\rule{5cm}{0.2mm}
%\psfig{figure=filename.ps,height=1.5in}
       \setlength{\unitlength}{1mm}
       \begin{picture}(80,60)
       \put(25,5){\epsfxsize=6cm \epsfbox{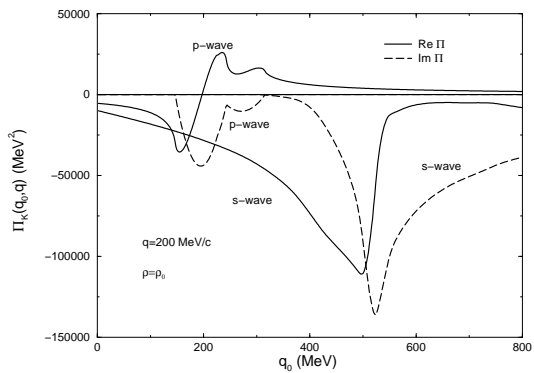}}
       \end{picture}
\caption{
Kaon self-energy.
\label{fig:mes98fig4}}
\end{figure}

\begin{figure}[ht]
%\rule{5cm}{0.2mm}\hfill\rule{5cm}{0.2mm}
%\vskip 5 cm
%\rule{5cm}{0.2mm}\hfill\rule{5cm}{0.2mm}
%\psfig{figure=filename.ps,height=1.5in}
       \setlength{\unitlength}{1mm}
       \begin{picture}(80,60)
       \put(25,5){\epsfxsize=6cm \epsfbox{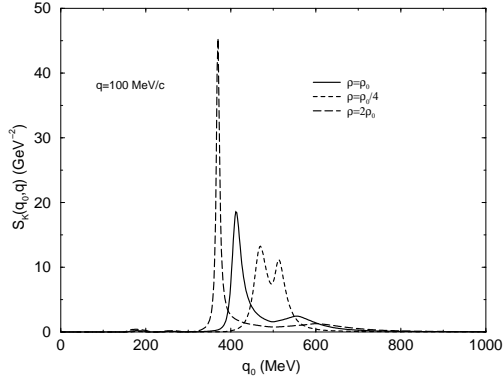}}
       \end{picture}
\caption{
$K^-$ spectral density for a momentum $q=100 MeV/c$ at several
densities: $\rho_0/4$ (short-dashed line), $\rho_0$ (solid line)
and $2\rho_0$ (long-dashed line).
\label{fig:mes98fig5}}
\end{figure}

The real (solid lines) and imaginary (dashed lines) parts of the
s- and p-wave $K^-$ self-energy at $\rho_0$ are shown in
Fig.~\ref{fig:mes98fig4} as functions of
$q_0$ for a $K^-$ momentum q$=200$ MeV/c. The p-wave
self-energy shows the typical shape of two Lindhard functions,
corresponding to the
$\Lambda$-- and $\Sigma$--hole excitations, and is located
at low $q_0$ values, where the contribution of the real part is
comparable to that of the
s-wave self-energy. However, the position of the new
$K^-$ pole is basically determined by the dominant s-wave
piece, especially for low momentum values. This can be seen
in Fig.~\ref{fig:mes98fig5}, where the
$K^-$ spectral density is plotted as a function of $q_0$ for a
momentum q$=100$ MeV/c and for several densities. The p-wave
self-energy
is responsible for the appearance of some extra strength,
barely seen on the scale of the figure,
to the left of the peaks.
At $\rho_0=\rho_0/4$ a two-mode excitation is clearly visible.
The left peak
corresponds to the $K^-$ pole branch, appearing at an energy
smaller
than $m_K$ due to the attractive medium effects, and the right
one
corresponds to the $\Lambda$(1405)-hole state, which is located
above $m_K$ due to Pauli blocking which
shifts the $\Lambda(1405)$ resonance to higher energies.
This
last excitation mode disappears as density increases, reflecting
the fact that the $\Lambda(1405)$ dissolves in dense
matter.

\begin{figure}[ht]
%\rule{5cm}{0.2mm}\hfill\rule{5cm}{0.2mm}
%\vskip 5 cm
%\rule{5cm}{0.2mm}\hfill\rule{5cm}{0.2mm}
%\psfig{figure=filename.ps,height=1.5in}
       \setlength{\unitlength}{1mm}
       \begin{picture}(80,60)
       \put(25,5){\epsfxsize=6cm \epsfbox{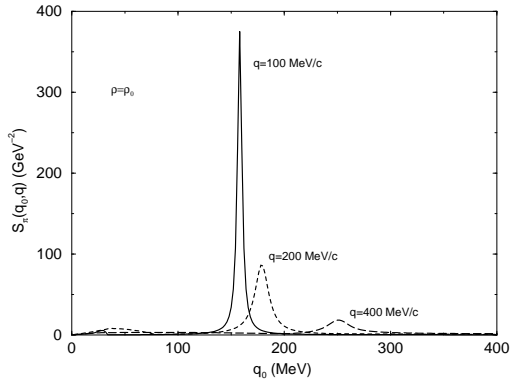}}
       \end{picture}
\caption{
Pion spectral density at $\rho_0$ for several momentum values:
100 (solid line), 200 (short-dashed line) and 400 (long-dashed
line) MeV/c.
\label{fig:mes98fig6}}
\end{figure}

The pion spectral density at $\rho_0$ is shown in
Fig.~\ref{fig:mes98fig6} as a
function of $q_0$ for several momentum values. Each peak
shows the position of the corresponding medium modified pion pole
and the
structures at the left show the strength due to
particle-hole excitations. It is clearly seen that the deviation
from
the free pion energy, $\sqrt{m_\pi^2 + {\vec q}\,^2}$, increases
as
the momentum increases and this moves the $\pi \Sigma$ and $\pi
\Lambda$ thresholds of the intermediate states in the loops to
much
lower energies.

The influence of the dressed mesons can be included in the
scheme by
replacing the free meson propagator
in Eq. (\ref{eq:gprop}) by that of
Eq. (\ref{eq:dressed}). Then, taking the Pauli
principle on the nucleons also into account, the loop integral
becomes
\begin{eqnarray}
G_l(\sqrt{s},{\vec P},\rho)&= &\int_0^\infty d\omega \int_{\mid
{\vec
q} \mid < q_{max}} \frac{d^3 q}{(2 \pi)^3}
\frac{M_l}{E_l (\vec{q}\,)}  S_l(\omega,{\vec q},\rho) \nonumber
\\
& x & \left\{
\frac{\theta(\mid {\vec P}-{\vec q} \mid - p_F)}{\sqrt{s}- \omega
- E_l (\vec{q}\,)
+ i \epsilon} +
\frac{\theta( p_F - \mid {\vec P}-{\vec q} \mid)}
{\sqrt{s}- \omega - E_ (\vec{q}\,) - i
\epsilon} \right\} \ .
\label{eq:gpropd}
\end{eqnarray}

When the dressing of the $K^-$ is incorporated,
the problem needs to be solved self-consistently since
the loop integral of Eq.~(\ref{eq:gpropd}), required to obtain
the in-medium ${\bar K}N$ interaction, $T^{\rm eff}$,
and thus the $K^-$ self-energy through Eq. (\ref{eq:selfka}),
also depends on the $K^-$ propagator. A self-consistent
calculation
has been attempted in Ref.~\cite{Lutz98}, where it is found that
the
repulsive shift on the $\Lambda(1405)$ mass induced by Pauli
blocking
is compensated by the use of an attractive $K^-$ self-energy
in the intermediate states. The mass of the resonance turns out to
be
close to its free space value and the width becomes larger as
density increases.

Work is in progress to check these predictions
within our model,
which also attempts at incorporating the strong medium
modifications
of the pion.

\section{Conclusions}

We have studied the s-wave ${\bar K}N$ interaction on the basis
of a coupled-channel Lippmann Schwinger equation, using the lowest
order chiral Lagrangian and one cut-off parameter. This simple
model gives an excellent agreement with all the low energy
scattering data and the role of the $\eta$
intermediate channels, omitted in previous
investigations, has been found crucial.

We apply this model to the study of the $K^-$ properties in the
nuclear medium. Pauli blocking on the intermediate nucleon states
produces a positive shift on the mass of the $\Lambda(1405)$,
which
ends dissolving in the medium at densities close to normal nuclear
matter density.
The $K^-$ optical potential is around $-100$ MeV, half of what is
obtained from $K^-$ atom data analyses.

Both the $K^-$ self-energy, to which we also add a small p-wave from
hyperon particle-nucleon hole excitations, and the strongly
attractive pion self-energy need to be incorporated
self-consistently in the loops before any conclusion on the
in-medium $\Lambda$(1405) properties can be drawn.

\section*{Acknowledgments}

This work is partially supported
by DGICYT contract numbers PB95-1249 and PB96-0753,
 and by the
EEC-TMR Program under contract No. CT98-0169.

%\section*{Appendix}
%We can insert an appendix here and place equations so that they
%are given numbers such as Eq.~(\ref{eq:app}).
%\be
%x = y.
%\label{eq:app}
%\ee

\end{document}